\documentclass[aps,prl,showkeys,twocolumn,superscriptaddress,groupedaddress]{revtex4}  % for review and submission
\usepackage{graphicx}  % needed for figures
\usepackage{dcolumn}   % needed for some tables
\usepackage{bm}        % for math
\usepackage{amssymb}   % for math
\usepackage{latexsym,bm}
\usepackage[tbtags]{amsmath}
\usepackage{anyfontsize}
\usepackage{color}

\usepackage[all,cmtip]{xy}

\usepackage{rotating}
\usepackage{marginnote}

\usepackage[utf8]{inputenc}
\usepackage[english]{babel}

\usepackage{hyperref}
\hypersetup{
    colorlinks=true,
    linkcolor=blue,
    filecolor=magenta,
    urlcolor=cyan,
}

\usepackage{color}

\usepackage{braket}

\newtheorem{theorem}{Theorem}

\newtheorem{lemma}[theorem]{Lemma}

\usepackage{hyperref}
\hypersetup{
    colorlinks=true,
    linkcolor=blue,
    filecolor=magenta,
    urlcolor=cyan,
}

\begin{document}

% The following information is for internal review, please remove them for submission
\widetext
%\leftline{Scientific Report Ref: CPUT/SUN/2017-02-02}
%\leftline{Primary authors: Professor Bohua Sun}
%\leftline{To be submitted to (PRL, PRD-RC, PRD, PLB; choose one.)}
%\leftline{Comment to {\tt d0-run2eb-nnn@fnal.gov} by xxx, yyy}
%\centerline{\em D\O\ INTERNAL DOCUMENT -- NOT FOR PUBLIC DISTRIBUTION}

% the following line is for submission, including submission to the arXiv!!
%\hspace{5.2in} \mbox{Fermilab-Pub-04/xxx-E}

\title{On Plastic Dislocation Density Tensor}% Force line breaks with \\
%\thanks{Arguments in favor of understanding the universality of turbulent boundary layer}
%\input author_list.tex       % D0 authors (remove the first 3 lines
                             % of this file prior to submission, they
                             % contain a time stamp for the authorlist)
                             % (includes institutions and visitors)

\author{Bohua Sun}

\affiliation{Institute of Mechanics and Technology and College of Civil Engineering\\Xi'an University of Architecture and Technology, City of Xi'an 710055, China\\sunbohua@xauat.edu.cn}%

%\email{sunbohua@xauat.edu.cn}

%\date{\today}

\begin{abstract}
%\normalsize
This article attempts to clarify an issue regarding the proper definition of plastic dislocation density tensor. This study shows that the Ortiz's and Berdichevsky's plastic dislocation density tensors are equivalent with each other, but not with Kondo's one. To fix the problem, we propose a modified version of Kondo's plastic dislocation density tensor.
\end{abstract}

%Use showkeys class option if keyword
\maketitle

%\marginpar{
%\begin{turn}{90}
%\huge{\url{http://www.sharelatex.com}}
% Prof. Bohua Sun \url{sunbohua@yahoo.com}
%\end{turn}}

%\tableofcontents

%\clearpage

%\section{Plastic dislocation density tensor}
Plastic deformation is everywhere, from bending a fork to panel beating a car body. It is easy to be intrigued by a subject that pervades so many aspects of peoples' daily lives.

G.I. Taylor \cite{taylor1934} realized that plastic deformation could be explained in terms of the theory of dislocations, even since this view has become a consensus that mechanism of plastic deformation is the result of dislocation accumulation \cite{taylor1934,reina2011,reina2014,reina2016,reina2017,kondo1955a,kroner1955,
kroner1960,bilby1957,bilby,ortiz,sedov,cairola,aifantis,le1995,le1996a,le1996,le2014,le2015,sun2014,sun2016,sun2018}.
Accordingly, some plastic dislocation density tensors have been proposed \cite{kondo1955a,kroner1955,
kroner1960,bilby1957,bilby,ortiz,sedov}. However, they are totally different from each other and no any consensus in terms of definition of the plastic dislocation density tensor. The majority of past and contemporary authors following the original idea of Kondo \cite{kondo1955a,kroner1955,
kroner1960} and Bilby et al. \cite{bilby}, adopted the following definition of the resultant Burgers vector $ \bm b_{\mathrm{Kondo}}=\bm F^{e}\oint_c \bm F^{e-1}\cdot d \bm x=-\bm F^{e}\iint (\bm F^{e-1}\times \bm \nabla) \cdot d \bm A$, where $c$ is any close contour in the current configuration. Ortiz and Repetto \cite{ortiz} defined the resultant Burgers vector in a completely different way $ \bm b_{\mathrm{Ortiz}}=\oint_c \bm F^{p}\cdot d \bm x=-\iint \bm F^{p} \times \bm \nabla \cdot d \bm A =\iint \bm T_{\mathrm{Ortiz}} \cdot d \bm A$. Reina \emph{et al}. \cite{reina2011,reina2014,reina2016,reina2017} did a comprehensive and in depth studies on the Ortiz's definition $\bm T_{\mathrm{Ortiz}}=-\bm F^{p} \times \bm \nabla$. Berdichevsky \cite{sedov} introduced a measure of the resultant closure failure leading to the dislocation density tensor $\bm T_{\mathrm{Berdichevsky}}=-\bm F^{p-1}\cdot (\bm F^{p} \times \bm \nabla)$. Le \emph{et al.} \cite{le1995,le1996a,le1996,le2014,le2015} recommended to use the Berdichevsky's definition.

It is clear that a unification of the definition for plastic dislocation density tensor is still an issue. which the above-mentioned definition is the proper one? What is the relationship between those definitions? If the definition is not well defined, how to fix it?

\begin{figure}[b]
\centerline{\includegraphics[scale=0.7]{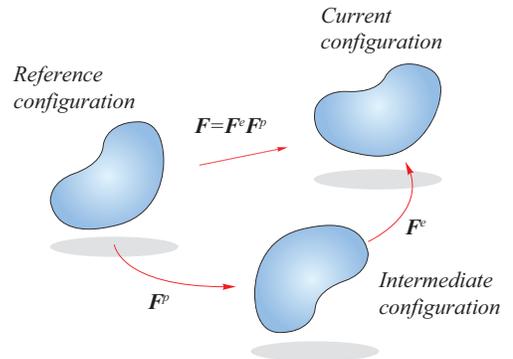}}
\caption{\label{fig1}Elasto-plastic deformation configuration: elastic deformation $\bm F^e$, plastic deformation $\bm F^p$ and total deformation $\bm F$}
\end{figure}

Phenomenologically the total elasto-plastic deformation gradient $\bm F=\bm g_i\otimes \bm G^i=F_{ij}\bm g^i\otimes\bm G^j=F_{ij}\bm g^i\bm G^j$ can be decomposed into the multiplication of elastic gradient namely $\bm F=\bm F^e \bm F^p$, is due to Bilby \emph{et. al.} \cite{bilby1957}, Kr\"oner \cite{kroner1960}, Lee and Liu \cite{leeliu}, and Lee \cite{ehlee}. The elastic deformation gradient $\bm F^e=\bm g_i\otimes \bm e^i=F^e_{ij}\bm g^i\otimes\bm e^j=F^e_{ij}\bm g^i\bm e^j$ and plastic gradient $\bm F^p=\bm e_i\otimes \bm G^i=F^p_{ij}\bm e^i\otimes\bm G^j=F^p_{ij}\bm e^i\bm G^j$,  $\bm G_i,\,\bm e_i,\,\bm g_i$ are the base vectors corresponding to the reference, intermediate and current configuration, respectively. The deformation decomposition is shown in Figure \ref{fig1}.

It should be noted that the elastic deformation $\bm F^e$ and plastic deformation $\bm F^p$ cannot be gradients of global maps, they are therefore called incompatible, namely $\bm F^e \times \bm \nabla\neq \bm 0$ and $\bm F^p \times \bm \nabla\neq \bm 0$ as well, where the operator $\bm \nabla=\bm G^k \nabla_k$ is gradient operator, and $\nabla_k$ is covariant derivative in reference configuration. Nevertheless, both $\bm F^e$ and $\bm F^p$ are orientation preserving so that $J_p=\det(\bm F^p)> 0$ and $J_e=\det(\bm F^e)> 0$. This means, $\bm F^p$ and $\bm F^e$ have inverse deformations, denoted correspondingly by $(\bm F^p)^{-1}$ and $(\bm F^e)^{-1}$.

In this short article, we will show that the Ortiz's and Berdichevsky's plastic dislocation density tensor are equivalent, while not equivalent with Kondo's one. To fix Kondo's problem, we can change Kondo's definition to following form
\begin{equation}\label{mk1}
\begin{split}
  \bm b_{\mathrm{Modified-Kondo}}&=\oint_c\cdot \bm F^{e-1}\cdot d \bm x\\
  &=-\iint (\bm F^{e-1}\times \bm \nabla) \cdot d \bm A\\
  &=\iint \bm T_{\mathrm{Modified-Kondo}} \cdot d \bm A,
\end{split}
\end{equation}
Thus, we have modified Kondo's plastic dislocation density tensor as follows
\begin{equation}\label{mk2}
  \bm T_{\mathrm{Modified-Kondo}}=-\bm F^{e-1}\times \bm \nabla.
\end{equation}
With this modified definition, later we will show that the modified Kondo's definition can be equivalent with both Ortiz's and Berdichevsky's plastic dislocation density tensor. To verify these, we need to prove a tensor identity at first.
\begin{lemma} Giving two 2nd order tensors $\bm A=\bm g_i\otimes \bm e^i=A_{ij}\bm g^i\otimes \bm e^j=A_{ij}\bm g^i\bm e^j$ and $\bm B=\bm e_i\otimes \bm G^i=B_{ij}\bm e^i\otimes \bm G^j=B_{ij}\bm e^i\bm G^j$, then we have tensor identity: $(\bm A\cdot \bm B)\times \bm \nabla=\bm A\cdot ( \bm B\times \bm \nabla)+(\bm A \times \bm \nabla)\cdot \bm B$.
\end{lemma}
\textbf{Proof}
\begin{equation}\label{eq1}
\begin{split}
  &(\bm A\cdot \bm B)\times \bm \nabla =(\bm A\cdot \bm B)\times \bm G^m\nabla_m\\
  &=[(\bm A\nabla_m) \cdot \bm B +\bm A\cdot (\bm B\nabla_m)]\times \bm G^m\\
  &=\bm A\cdot (\bm B\nabla_m)\times \bm G^m+(\bm A\nabla_m) \cdot \bm B \times \bm G^m,
  \end{split}
\end{equation}
where $\bm A\cdot (\bm B\nabla_m)\times \bm G^m=\bm A\cdot (\bm B \times \bm \nabla)$, and
\begin{equation}\label{eq2}
\begin{split}
  &(\bm A\nabla_m) \cdot \bm B \times \bm G^m\\
  &=[(\bm A  \bm \nabla)\cdot \bm G_m]\cdot [B_{kl}\bm e^k \bm G^l\times \bm G^m ]\\
   &=\bm A \bm \nabla \cdot B_{kl} (\bm e^k\cdot \bm G_m) (\bm G^l\times \bm G^m )\\
  &=\bm A \bm \nabla \cdot B_{kl} (\bm e_k\bm G^l)\tiny{\begin{array}{c}
   \bullet \\
   \times \\
   \end{array}}( \bm G_m \bm G^m )
  =\bm A  \bm \nabla \cdot \bm B \tiny{\begin{array}{c}
   \bullet \\
    \times \\
    \end{array}}\bm I\\
&=-\bm A  \bm \nabla \tiny{\begin{array}{c}
  \bullet \\
   \times \\
    \end{array}}\bm B
=-\bm A  \bm \nabla \tiny{\begin{array}{c}
  \bullet \\
   \times \\
    \end{array}}(\bm I\cdot \bm B)\\
&=-(\bm A  \bm \nabla \tiny{\begin{array}{c}
  \bullet \\
    \times \\
     \end{array}}\bm I)\cdot \bm B=(\bm A  \times \bm \nabla) \cdot \bm B,
  \end{split}
\end{equation}
where the unit tensor $\bm I=\bm G_m\bm G^m=\delta_{ij} \bm G^i\bm G^j$ in reference configuration. Therefore, we have proven the tensor identity, which has never been seen in literature.

Despite the incompatibility of elastic and plastic deformation, namely, $\bm F^e$ and $\bm F^p$, the total deformation $\bm F$ is compatible, it means that the total deformation must be gradient of global maps, thus it must satisfy compatible condition \cite{sun2016}, namely, the incompatible tensor $Inc(\bm F)=\bm F\times \bm \nabla=\bm 0$, which leads to $Inc(\bm F)=(\bm F^e \cdot \bm F^p)\times \bm \nabla=\bm 0$.

Applying the identity of tensor proved in the Lemma, we have
\begin{equation}\label{u1}
\begin{split}
&Inc(\bm F)=\bm F\times \bm \nabla=(\bm F^e \cdot \bm F^p)\times \bm \nabla\\
&=\bm F^e \cdot (\bm F^p \times \bm \nabla)+(\bm F^e \times \bm \nabla)\cdot \bm F^p= \bm 0.
 \end{split}
\end{equation}
Using the previous definitions of plastic dislocation density tensor, the above expression can be rewritten as
\begin{equation}\label{u2}
\begin{split}
&Inc(\bm F)=\bm F^e \cdot (-\bm T_{\mathrm{Oritz}})+(-\bm T_{\mathrm{Modified-Kondo}})\cdot \bm F^p\\
 &=\bm F\cdot (-\bm T_{\mathrm{Berdichevsky}})+(-\bm T_{\mathrm{Modified-Kondo}})\cdot \bm F^{p}\\
 &= \bm 0.
 \end{split}
\end{equation}
Therefore, we have their relationships:
\begin{align}
  \bm F^e \cdot \bm T_{\mathrm{Oritz}}+\bm T_{\mathrm{Modified-Kondo}}\cdot \bm F^p&=\bm 0, \label{p1}\\
  \bm F\cdot \bm T_{\mathrm{Berdichevsky}}+\bm T_{\mathrm{Modified-Kondo}}\cdot \bm F^{p}&=\bm 0,\label{p2}\\
  \bm F\cdot \bm T_{\mathrm{Berdichevsky}}-\bm F^e \cdot \bm T_{\mathrm{Oritz}}&=\bm 0.\label{p3}
\end{align}
Clearly the relations \ref{p1},\ref{p2} and \ref{p3} reveal that three definition of the plastic dislocation tensity tensor are equivalent.

In summary, this study shows that both Ortiz's and Berdichevsky's plastic dislocation density tensors are equivalent, and are proper definition. Although Kondo's definition is not proper one, it can be fixed by the modified version in Eq. \ref{mk2}.


\nocite{*}

\begin{thebibliography}{99}

\bibitem{taylor1934}
Taylor, G.I., The mechanism of plastic deformation of crystals. Part I. Theoretical. \emph{Proc. of the Royal Society of London}, Series A \textbf{145}(855) (1934)362-387

%\bibitem{nye} Nye, J.F., Some geometrical relations in dislocated crystals. \emph{Acta Metallurgica}, \textbf{1}(2) (1953)153-162

\bibitem{kondo1955a}
Kondo, K., Geometry of elastic deformation and incompatibility. \emph{Memoirs of the Unifying Study of the Basic Problems in Engineering Science by Means of Geometry}, (K. Kondo, ed.), \textbf{1}, Division C, Gakujutsu Bunken Fukyo-Kai, (1955)5-17

%\bibitem{kondo1955b} Kondo, K., Non-Riemannien geometry of imperfect crystals from a macroscopic viewpoint. \emph{Memoirs of the Unifying Study of the Basic Problems in Engineering Science by Means of Geometry}, (K. Kondo, ed.), \textbf{1}, Division D-I, Gakujutsu Bunken Fukyo-Kai, (1955) 6-17

\bibitem{bilby1957}
Bilby, B.A., Gardner, L.R.T., Stroh, A.N., Continuous distributions of dislocations and the theory of plasticity. Proceedings of the 9th International Congress of Applied Mechanics,  vol.8: 35-44, Universit\'e de Bruxelles, 1957

\bibitem{bilby}
Bilby, B. A., Bullough, R. and Smith, E., Continuous distributions of dislocations: a new application of the methods of non-Riemannian geometry. \emph{Proc. of the Royal Society of London}, A \textbf{231}(1185) (1995)263-273

\bibitem{kroner1955}
Kr\"oner, E., Das Fundamentalintegral der anisotropen elastischen Versetzungsdichte und Spannungsfunktionen. \emph{Z. Phys.}, \textbf{142} (1955)463-475

\bibitem{kroner1960}
Kr\"oner, E., Allgemeine kontinuumstheorie der versetzungen und eigenspannungen. \emph{Arch. Rational Mech. Anal.}, \textbf{4}(4) (1960)273-334

\bibitem{ortiz}
Ortiz, M., Repetto, E.A., Noncovex energy minimization and dislocation structures in ductile single crystals. J.Mech.Phys.Solids 47 (1999)397-462

\bibitem{sedov}
Sedov, L.I. and Berdichevsky, V.L., A Dynamic theory of continual dislocations. In \emph{Mechanics of generalized continua} (ed. E. Kr\"oner), 215-238. Springer-Verlag, Berlin, 1967.

\bibitem{cairola}
Gairola, B.K.D., Chapter 4: Nonlinear elastic problems. \emph{Dislocations in Solids}, Edited by F.R.N. Nabarro, North-Holland Publishing Co. (1979).

\bibitem{aifantis}
Aifantis, E.C. The physics of plastic deformation. \emph{Int. J. of Plasticity}, \textbf{3} (1987)211-247

\bibitem{reina2011}
Conti, S., Dolzmann, G., Kreisbeck, C., Asymptotic behavior of crystal plasticity with one slip system in the limit of rigid elasticity. \emph{SIAM J. Math. Anal.}
43 (2011)2337-2353

\bibitem{reina2014}
Reina, C., Conti, S., Kinematic description of crystal plasticity in the finite kinematic framework: a micromechanical understanding of $\bm F=\bm F^e\bm F^p$. \emph{J. Mech. Phys. Solids} 67 (2014)40-61

\bibitem{reina2016}
Reina, C., Schl\"omerkemper, A., Conti, S., Derivation of $\bm F=\bm F^e\bm F^p$ as the continuum limit of crystalline slip. \emph{J. Mech. Phys. Solids} 89 (2016)231-254

\bibitem{reina2017}Reina, C. and Conti, S. Incompressible inelasticity as an essential ingredient for the validity of the kinematic decomposition. \emph{J. Mech. Phys. Solids} 107 (2017)322-342

\bibitem{le1995}Le, K.C. and Stumpf, H. Nonlinear continuum theory of dislocations. \emph{Int. J. of Engineering Science}, \textbf{34} (1995)339-358

\bibitem{le1996a}Le, K.C. and Stumpf, H. On the determination of the crystal reference in nonlinear continuum theory of dislocations. \emph{Proc. Roy. Soc. London}, A \textbf{452},(1996) 359-371

\bibitem{le1996}Le, K.C. and Stumpf, H., A model of elastoplastic bodies with continuously distributed dislocations. \emph{Int J. of Plasticity}, \textbf{12} (1996)611-627

\bibitem{le2014}
Le, K.C. and G\"unther, C. Nonlinear continuum dislocation theory revisited. \emph{Int. J. of Plasticity}, \textbf{53} (2014)164-178

\bibitem{le2015}
Le, K.C., Three-dimensional continuum dislocation theory. \emph{Int. J. of Plasticity}, \textbf{76}, (2016) 213-230

\bibitem{leeliu}
Lee, E.H. and Liu, D.T., Elasticplastic theory with application to plane-wave analysis. \emph{J. Appl. Phys.}, 38(1967):19-27

\bibitem{ehlee}
Lee, E.H., Elastic-plastic deformation at finite strains. \emph{J. Appl. Mech.}, \textbf{36}(1)(1969) 1-6

\bibitem{sun2014}
Sun, B. Rational interpretation of the postulates in plasticity. \emph{Results in Physics} \textbf{4} (2014) 10-11

\bibitem{sun2016}
Sun, B. Incompatible deformation field and Riemann curvature tensor. \emph{Appl. Math. Mech.}, 38(3) (2017)311-332

\bibitem{sun2018}
Sun, B. On plastic dislocation density tensor.\emph{Preprints} 2018, 2018100330 (doi: 10.20944/preprints201810.0330.v1)


\end{thebibliography}
\end{document}